\documentclass[aps,draft]{revtex4}
\usepackage[dvips]{graphicx}
\begin{document}

\title{ Nonlinear excitation of photonikos and plasmons by high-power, short pulse lasers }
\author{Levan N.Tsintsadze}
\thanks{Also at Department of Plasma Physics, E.Andronikashvii Institute of Physics, Tbilisi,
Georgia}
\author{Hitoshi Hojo}
\affiliation{Graduate School of Pure and Applied Sciences, University of Tsukuba, Tsukuba, Japan}

\date{\today}

\begin{abstract}
Modulational excitation of longitudinal photons (photonikos) and electron Langmuir waves, as well as ion sound waves by an incoherent strong and superstrong radiation (high-power short pulse lasers, non-thermal equilibrium cosmic field radiation, etc.) in plasmas are investigated. A simultaneous generation of photonikos and plasmons are demonstrated. Furthermore, the kinetic instability is considered when a low frequency photonikos are generated alone. Growth rates of these new modes are obtained.
\end{abstract}

\maketitle

Electromagnetic (EM) radiation in plasma is a fundamental physical system which has played a crucial role in opening up new frontiers in physics, such as the fast ignition in laser fusion \cite{tab}, plasma based high-energy particle acceleration
\cite{tsin74},\cite{taj}, the electron-positron and the neutrino-antineutrino pair production \cite{lif}, \cite{zelrel}, \cite{she}, \cite{sna}, nonlinear optics \cite{akhm}, optically induced nuclear fission \cite{boy}, etc. One of the most salient phenomenon in the above mentioned problems of laboratory plasmas, as well as astrophysical and space plasmas are the relativistic parametric and modulational instabilities. Which gained prominence largely due to its importance in connection with the strongly nonlinear structures in plasmas, the deposition of EM energy through different modes into the plasma, heating and/or acceleration of plasma particles to very high energies. The relativistic parametric instabilities has a long history starting with the work of Tsintsadze \cite{tsin70}. Since then a great number of theoretical and simulation results were reported \cite{max}. Recent progress in development of high-power, short pulse lasers has renewed the interest in this phenomenon \cite{ltsin91},\cite{mor94},\cite{tsin94},\cite{gue95}. The above treatments where restricted to the case of monochromatic EM waves. For ultrashort pulses the bandwith of coherent wave is increasingly broad. Even if the bandwith may be initially narrow, its spectrum may eventually broaden, either as a result of several kinds of instability processes, or as the result of other nonlinear wave-wave interaction processes.

In order to study the interaction of spectrally broad relativistically intense EM waves with a plasma, in previous papers \cite{ltsin98}, \cite{ntsin98} starting from the fully relativistic equations, we have derived a general kinetic equation for the photon gas incorporating two forces of distinct nature. One force appears due to the redistribution of electrons in space, $\nabla n_e$, and time, $\frac{\partial n_e}{\partial t}$. The other force arises by the variation of the shape of wavepacket. In other words, this force originates from the alteration of the average kinetic energy of the electron oscillating in a rapidly varying field of EM waves, and is proportional to $\nabla\frac{1}{\gamma}$ and $\frac{\partial}{\partial t}\frac{1}{\gamma}$, where $\gamma$ is the relativistic Lorentz factor.

In the field of superstrong femtosecond pulses, it is expected that the character of the nonlinear response of medium would
radically change. At high intensities the motion of free electrons near the focal volume would be extremely relativistic. Thus, the relativistic nonlinear effect, which is basically associated with the increase in the electron mass, will tend to determine the
dynamics of EM pulses. Currently, lasers produce pulses whose intensity approaches $10^{22} W/cm^2$ \cite{mour}. At these
intensities, the highly nonlinear and complicated relativistic dynamics of the laser-plasma system gives rise to a number of
interesting phenomena, for example the Bose-Einstein condensation (BEC) and a new intermediate state of the photon gas
\cite{ltsin02}. It was shown in Refs.\cite{ltsin96} that the behavior of photons in a plasma is radically different from the
one in a vacuum. Namely, plasma particles perform oscillatory motion in the field of EM waves affecting the radiation field.
Photons acquire the rest mass, $m_\gamma$, and become one of the Bosons in plasmas and posses all characteristics of nonzero rest
mass, i.e., we may say that the photon is the elementary particle of the optical field. This difference leads to certain novel
phenomenon, such as photonikos \cite{ltsin03}, \cite{ltsin07}, \cite{ntsin07}, which originates from the decay process. Namely,
the photon passing through the photon bunch absorbs and emits photonikos, with frequencies $\Omega=\mp(\omega-\omega\prime)$ and
wavevectors $\vec{q}=\mp(\vec{k}-\vec{k}\prime)$. In Refs.\cite{ltsin02},\cite{ltsin03} we have shown
that under certain conditions the photon-photon interactions dominate the photon-plasma particle interactions. So that under
such conditions the variation of the plasma density can be neglected in comparison with the variation of the photon density.
In recent 2 dimensional fully relativistic particle-in-cell simulations (Fourier space code - EPIC3D-AP) \cite{ltsin08}, which is based on potential form,
the simultaneous  emission of photonikos and plasmons by the laser pulse were observed, as well as the low-frequency photons that can be associated with BEC photons were seen. Thus, it is of interest to examine these new modes.

In this letter, we study analytically the nonlinear excitation of photonikos and plasmons by high-power, short pulse lasers. As we will see in the following there are cases, when both photonikos and plasmons are simultaneously generated, that confirm the results of recent simulations \cite{ltsin08}. For our purpose, we employ the general dispersion relation obtained in Refs.\cite{ltsin98}, \cite{ntsin98}, which reads

\begin{eqnarray}
\label{startdisp}
\varepsilon\Bigl(1+\frac{\omega_{Le}^2}{2\gamma_{\circ}^{2}}\int\frac{d{\bf
k}}{ (2\pi)^{3}}\frac{ {\bf P}_{\circ}^{+}-{\bf
P}_{\circ}^{-}}{{\bf q}{\bf k}c^2-\Omega\omega(k)} \Bigr)+
(1+\delta\varepsilon_{i})\delta\varepsilon_{e}\frac{q^{2}c^{2}}
{2\gamma_{\circ}^{2}}\int\frac{d{\bf k}}{ (2\pi)^{3}}\frac{ {\bf
P}_{\circ}^{+}-{\bf P}_{\circ}^{-}}{{\bf q}{\bf
k}c^2-\Omega\omega(k)}=0 \ ,
\end{eqnarray}
where
\begin{eqnarray}
\label{definit}
\varepsilon=1+\delta\varepsilon_{e}+\delta\varepsilon_{i},
\hspace{.4cm} \delta\varepsilon_{\alpha}=\frac{4\pi
e^{2}}{q^{2}}\int\frac{({\bf q}
\partial f_{\circ\alpha}/\partial {\bf p})}{\Omega-{\bf q}{\bf v_\alpha}}d{\bf p},
\hspace{.3cm} \omega_{L\alpha}^2=\frac{4\pi e^2n_{0e}}{m_{0\alpha}\gamma_0}\ , \\
\nonumber \gamma_0=\sqrt{1+Q_0}=\sqrt{1+2\int\frac{d{\bf
k}}{(2\pi)^3}P_0} \hspace{.3cm} and  \hspace{.3cm}
P_0=\frac{e^2A(k)A^*(k)}{(m_{0e}c^2)^2} \hspace{.3cm} is \ the \
spectral \ function\ ,
\end{eqnarray}
Eq.(\ref{startdisp}) rewrite as
\begin{eqnarray}
\label{refirst}
\varepsilon\Bigl(1+\frac{1}{A}\Bigr)+(1+\delta\varepsilon_{i})
\delta\varepsilon_{e}\frac{q^{2}c^{2}}{\omega_{Le}^2}=0 \ ,
\end{eqnarray}
where
\begin{eqnarray}
\label{Adefin}
A=\frac{\omega_{Le}^2}{2\gamma_{\circ}^{2}}\int\frac{d{\bf k}}{
(2\pi)^{3}}\frac{ {\bf P}_{\circ}^{+}-{\bf P}_{\circ}^{-}}{{\bf
q}{\bf k}c^2-\Omega\omega(k)}=
\frac{\omega_{Le}^2}{2\gamma_{\circ}^{2}}\int\frac{d{\bf k}}{
(2\pi)^{3}}P_0\Bigl\{\frac{1}{\Omega-{\bf q}{\bf
u}+q^2c^2/2\omega}-\frac{1}{\Omega-{\bf q}{\bf u}-q^2c^2/2\omega}-
\\ \nonumber
\imath\pi\Bigl(\delta(\Omega-{\bf q}{\bf
u}+\frac{q^2c^2}{2\omega})-\delta (\Omega-{\bf q}{\bf
u}-\frac{q^2c^2}{2\omega})\Bigr)\Bigr\}
 \ .
\end{eqnarray}
If $A\gg 1$, then Eq.(\ref{refirst}) reduces to Eq.(20) of
Ref.\cite{ltsin98}.

For the photon gas we use the spectral Gaussian distribution
function
\begin{eqnarray}
\label{sgauss} P_0=Q_0(2\pi\sigma_k^2)^{-3/2}exp\Bigl(-\frac{({\bf
k}-{\bf k}_0)^2}{ 2\sigma_k^2}\Bigr) \ ,
\end{eqnarray}
where
\begin{eqnarray*}
\label{q0} Q_0=\frac{e^2\mid
A(k_0)\mid^2}{(m_{0e}c^2)^2}=\gamma_0^2-1 \ .
\end{eqnarray*}
Note that if there is no variation of the plasma density $\delta
n_\alpha=0,$ $\delta\varepsilon_\alpha=0$, then we have the equation
that was studied in Ref.\cite{ltsin03}.

We now rewrite Eq.(\ref{refirst}) taking into account poles in the
integrals
\begin{eqnarray}
\label{poleA} \Omega-{\bf q}{\bf u}=0 \ ,
\end{eqnarray}
\begin{eqnarray}
\label{poledeps} \Omega-{\bf q}{\bf v_\alpha}=0 \ ,
\end{eqnarray}
where
\begin{eqnarray*}
\label{usgansaz} {\bf u}=\frac{{\bf k}c^2}{\omega(k)}\ .
\end{eqnarray*}
Using the well known relation
\begin{eqnarray}
\lim_{\varepsilon\rightarrow 0}\frac{1}{x+\imath\varepsilon}=\wp
\frac{1}{x}-\imath\pi\delta(x) \ , \label{wellk}
\end{eqnarray}
and recalling \cite{ltsin03}
\begin{eqnarray}
\label{recall} A=ReA+iImA=A_0+\imath A_1 \ , \\ \nonumber
\delta\varepsilon_\alpha=\delta\varepsilon_\alpha^\prime+\imath\delta\varepsilon_\alpha^{
\prime\prime}, \hspace{.3cm} and \hspace{.3cm}
\Omega=\Omega^\prime+\imath\Omega^{\prime\prime}
\end{eqnarray}
and assuming that $A_0\gg \mid A_1\mid$, and
$\mid\delta\varepsilon_\alpha^\prime\mid\gg\mid\delta\varepsilon_\alpha^{
\prime\prime}\mid $, we rewrite Eq.(\ref{refirst}) as
\begin{eqnarray}
\label{maineq}
\varepsilon\Bigl(1+\frac{1}{A_0}-\imath\frac{A_1}{A_0^2}\Bigr)+(1+\delta\varepsilon_{i})
\delta\varepsilon_{e}\frac{q^{2}c^{2}}{\omega_{Le}^2}=0 \ .
\end{eqnarray}

We first consider the excitation of the longitudinal photons and
electron Langmuir waves ($\Omega^\prime\sim\omega_{Le}$, and
$\Omega\gg\omega_{Li}$, or $\delta\varepsilon_{i}=0$). In this
case
\begin{eqnarray}
\label{epsilon}
\varepsilon=1+\delta\varepsilon_{e}^\prime+\imath\delta\varepsilon_{e}^{
\prime\prime}=1-\frac{\omega_{Le}^2(1+3q^2r_{De}^2)}{\Omega^2}-\imath\frac{
4\pi^2e^2m_{0e}}{q^2}\Bigl(\frac{\partial f_{0e}}{\partial
p_x}\Bigr)_{v_x=\Omega/q} \ ,
\end{eqnarray}
\begin{eqnarray}
\label{oneplus}
1+\frac{1}{A_0}=\frac{1}{q^2V_E^2}\Bigl\{(\Omega-{\bf q}{\bf
u})^2-q^2(V_s^2-V_E^2)-\alpha^2q^4\Bigr\} \ ,
\end{eqnarray}
\begin{eqnarray}
\label{Aone}
A_1=-\sqrt{\frac{\pi}{2}}\frac{\omega^3}{(\sigma_kc)^3}\frac{V_E^2}{c^2}\frac{
(\Omega-{\bf q}{\bf u})}{qc}exp\Bigl(-\frac{3}{2} \frac{
(\Omega-{\bf q}{\bf u})^2}{q^2V_s^2}\Bigr)\ ,
\end{eqnarray}
where
\begin{eqnarray}
\label{vevsdef}
V_E^2=\frac{c^2}{2}\Bigl(\frac{\omega_{Le}}{\omega}\Bigr)^2\frac{\gamma_0^2-1}{\gamma_0^2},
\hspace{.3cm} V_s=c\sqrt{3}\frac{\sigma_k c}{\omega(k_0)},
\hspace{.3cm} \alpha=\frac{c^2}{2\omega(k_0)}\ ,
\end{eqnarray}
and $r_{De}$ is the Debay length for electrons.

We now neglect the small imaginary term in Eq.(\ref{maineq}) and
examine the following dispersion relation
\begin{eqnarray}
\label{dispboth}
\varepsilon_e^\prime\Bigl(1+\frac{1}{A_0}\Bigr)+\frac{q^2c^2}{\omega_{Le}^2}
\delta\varepsilon_{e}^\prime=0 \ ,
\end{eqnarray}
or more explicitly
\begin{eqnarray}
\label{exdispboth}
\Bigl\{\Omega^2-\omega_{Le}^2(1+3q^2r_{De}^2)\Bigr\}\Bigl\{(\Omega-{\bf
q}{\bf u})^2-q^2U^2-\alpha^2q^4\Bigr\}=q^4V_E^2c^2(1+3q^2r_{De}^2)
\ ,
\end{eqnarray}
where
\begin{eqnarray*}
U^2=V_s^2-V_E^2, \hspace{.3cm} {\bf u}({\bf k}_0)=\frac{{\bf
k}_0c^2}{\omega(k_0)}, \hspace{.3cm}{\bf q}{\bf u}=qucos\Theta \ .
\end{eqnarray*}
The maximum growth rate we obtain if the density energy of photons
compensate the kinetic "thermal" energy of photons and diffraction
term, i.e.,
\begin{eqnarray}
\label{compes} V_s^2+\alpha^2q^2=V_E^2
\end{eqnarray}
and from Eq.(\ref{exdispboth}) for
$\Omega=\omega_{Le}(1+\frac{3}{2}q^2r_{De}^2)+\delta\approx{\bf
q}{\bf u}+\delta$, we obtain
\begin{eqnarray}
\label{delta3}
\delta^3=\frac{1}{2\omega_{Le}}q^4V_E^2c^2(1+3q^2r_{De}^2)^{1/2} \
,
\end{eqnarray}
or
\begin{eqnarray}
\label{grate1}
Im\delta=\frac{\sqrt{3}}{2}\Bigl(\frac{qc}{2\omega_{Le}}\frac{V_E^2}{c^2}
(1+3q^2r_{De}^2)^{1/2}\Bigr)^{1/3}qc \ .
\end{eqnarray}
If the relation (\ref{compes}) does not hold, then
Eq.(\ref{exdispboth}) has the unstable solution such as
\begin{eqnarray}
\label{unstsol} \Omega=\omega_{Le}\sqrt{1+3q^2r_{De}^2}+\delta
\hspace{.3cm} and \hspace{.3cm} \Omega={\bf q}{\bf
u}-q\sqrt{U^2+\alpha^2q^2}+\delta \ .
\end{eqnarray}
From Eq.(\ref{exdispboth}) follows for the imaginary part of
$\delta$
\begin{eqnarray}
\label{grate2} Im\delta=\frac{qV_E}{2\omega_{Le}}
qc\Bigl(\frac{
 {\bf q}{\bf u}}{\omega_{Le}\sqrt{1+3q^2r_{De}^2}}-1\Bigr)^{-1/2}\
.
\end{eqnarray}
Note that ${\bf q}{\bf u}>\omega_{Le}\sqrt{1+3q^2r_{De}^2}$
always, as follows from relation (\ref{unstsol}).
Eqs.(\ref{grate1}) and (\ref{grate2}) demonstrate the modulational
excitation of photonikos and plasmons simultaneously.

We now consider the kinetic instability for the case when
$q^2v_{tre}^2<\Omega^{\prime 2}<\omega_{Le}^2$, which means that
in this case the low frequency photonikos are generated alone.
From Eq.(\ref{maineq}) a simple calculation gives for the
imaginary part of $\Omega(q)$ the following expression
\begin{eqnarray}
\label{photalone}
\Omega^{\prime\prime}=-\sqrt{\frac{\pi}{8}}\Bigl\{(\Omega^\prime-{\bf
q}{\bf
u})\frac{\omega_{Le}}{\Omega^\prime}\Bigl(\frac{\omega_{Le}}{qc}\Bigr)^3
\Bigl(\frac{V_E}{c}\Bigr)^2\Bigl(\frac{\omega}{\sigma_kc}\Bigr)^3
exp\Bigl(-\frac{3}{2}\frac{(\Omega^\prime-{\bf q}{\bf
u})^2}{2q^2V_s^2}\Bigr)+\Omega^\prime\Bigl(\frac{\Omega^\prime}{
qv_{tre}}\Bigr)^3exp\Bigl(-\frac{\Omega^{\prime
2}}{2q^2v^2_{tre}}\Bigr)\Bigr\} \ .
\end{eqnarray}
From Eq.(\ref{photalone}) follows that $\Omega^{\prime\prime}$
changes sing and becomes positive if the Cherenkov condition is
satisfied
\begin{eqnarray}
\label{chercon}
u>\frac{\Omega^\prime}{qcos\Theta}\Bigl\{1+\frac{\Omega^\prime}{\omega_{Le}}\Bigl(\frac{
qc}{\omega_{Le}}\Bigr)^3
\Bigl(\frac{c}{V_E}\Bigr)^2\Bigl(\frac{\sigma_kc}{\omega}\Bigr)^3
exp\Bigl(-\frac{\Omega^{\prime 2}}{2q^2v^2_{tre}}\Bigr)\Bigr\} \ .
\end{eqnarray}
So that this instability leads to the excitation of photonikos.
Here we have supposed that $\mid\Omega^\prime-{\bf q}{\bf
u}\mid<qV_s$.

We next study the range of frequencies
\begin{eqnarray}
\label{ranfre} q^2v_{tri}<\Omega^\prime <q^2v_{tre}
\end{eqnarray}
and the case when $\delta n_i\neq0$. In this case
\begin{eqnarray}
\label{epsion}
\varepsilon=\varepsilon^\prime+\imath\delta\varepsilon^{
\prime\prime}=1+\frac{1}{q^2r_{De}^2}+\imath\sqrt{\frac{\pi}{2}}\frac{\omega_{Le}^2\Omega}{
(qv_{tre})^3}-\frac{\omega_{pi}^2}{\Omega^2}\ .
\end{eqnarray}

First, we neglect the small imaginary terms in Eqs.(\ref{epsion})
and (\ref{Adefin}), and use Eq.(\ref{maineq}) to obtain the
following dispersion relation
\begin{eqnarray}
\label{dision} (\Omega^2-\Omega_s^2)\Bigl\{(\Omega-{\bf q}{\bf
u})^2-q^2(U^2+\alpha^2q^2)\Bigr\}+\frac{q^4c^2V_E^2(\Omega^2-\omega_{pi}^2)}{
\omega_{Le}^2(1+q^2r_{De}^2)}=0 \ ,
\end{eqnarray}
where $\Omega_s=\frac{qv_s}{\sqrt{1+q^2r_{De}^2}}$ is the ion
sound frequency.

If the length of waves is shorter than the Debay length, i.e.,
$\lambda\ll r_{De}$, then $\Omega_s\rightarrow\omega_{pi}$, and
Eq.(\ref{dision}) reduces to
\begin{eqnarray}
\label{reddision} (\Omega-{\bf q}{\bf
u})^2-q^2(V_s^2+\alpha^2q^2)+q^2V_E^2\Bigl(1+\frac{c^2}{v_{tre}^2}\Bigr)=0
\ .
\end{eqnarray}
This relation demonstrates that the ions play no role in the
instabilities. Whereas, in the presence of the hot electrons, the
growth rate becomes large, because of the coupling term (the last
term in Eq.(\ref{dision})), as $c^2/v_{tre}^2$, and the imaginary
part of the frequency is
\begin{eqnarray*}
Im\Omega=\Omega^{\prime\prime}\simeq qV_E\frac{c}{v_{tre}}\ .
\end{eqnarray*}

In the opposite case, i.e., $q^2r_{De}^2\ll 1$, we have the
modulational excitation of photonikos and ion sound waves
simultaneously. That is the photon flux triggers the both waves.
To show this, we first consider the case when the condition
(\ref{compes}) is satisfied. In this case for  the
coincide roots $\Omega=\Omega_s+\delta={\bf q}{\bf u}+\delta$,
where $\Omega_s\gg\mid\delta\mid$, we obtain from
Eq.(\ref{dision})
\begin{eqnarray}
\label{grion1} Im\delta=\frac{\sqrt{3}}{2}
qc\Bigl(\frac{m_e}{2m_i}\frac{V_E^2}{cv_s}\Bigr)^{1/3} \ .
\end{eqnarray}
Next, if the relation (\ref{compes}) does not hold, then
Eq.(\ref{dision}) has the unstable solution such as
\begin{eqnarray}
\label{unsol} \Omega=\Omega_s+\delta \hspace{.3cm} and
\hspace{.3cm} \Omega={\bf q}{\bf u}-q\sqrt{U^2+\alpha^2q^2}+\delta
\end{eqnarray}
and we get the growth rate from Eq.(\ref{dision}) as
\begin{eqnarray}
\label{grion2}
Im\delta=\frac{qc}{2}\Bigl(\frac{V_E}{v_{tre}}\Bigr)
\frac{1}{(1+q^2r_{De}^2)^{1/2}}\frac{1}{({\bf q}{\bf
u}/\Omega_s-1)^{1/2}}\ .
\end{eqnarray}
Which is true for the Cherenkov condition $u >
\frac{\Omega_s}{qcos\Theta}=\frac{v_s}{cos\Theta}$. Furthermore, the equation
(\ref{grion2}) is valid when $V_E\ll v_{tre}$, since
$\Omega_s=qv_s>Im\delta$.

To summarize, we have shown a simultaneous nonlinear excitation of photonikos and plasmons in the interaction of relativistically intense nonmonochromatic radiation bunches with a nonmagnetized plasma. Cases without and with ion dynamics are discussed. The generation of only low frequency photonikos is also confirmed. The growth rates of these new modes, which have no counterpart in the case of monochromatic EM waves, are obtained. These investigations may play an essential role in advanced fusion concepts and advanced accelerators, as well as for the description of extremely complex phenomena that appear only in energetic astrophysical systems and in experiments modeling the high energy density astrophysics in the laboratory. It was argued in Ref.\cite{okun} that the $m_\gamma$ would lead to a catastrophic emission of longitudinal photons. This gas may play the decisive role in the expansion of the Universe. In addition in the fireball model of Gamma-ray bursts, the afterglow may be due to the decay process discussed in this and the previous papers \cite{ltsin02}, \cite{ltsin03}. A cursory examination of burst profiles indicates that some are chaotic and spiky with large fluctuations on all time scales, while others show rather simple structures with few peaks. However, some bursts are seen with both characteristics present within the same burst \cite{fis}.

\end{document}